%% file: Arxiv.tex
\documentclass[twocolumn,showpacs,superscriptaddress,email,floatfix,longbibliography,prl]{revtex4-2}
\usepackage{braket}
\usepackage{natbib}
\usepackage{xcolor}
\usepackage{amsmath}
\usepackage{array}
\usepackage{tabularx}
\usepackage{mathtools}
\usepackage{commath}
\usepackage{bbold}
\usepackage{graphicx}
\usepackage{amsmath}
\usepackage{lipsum}
\graphicspath{ {Images/} }
\usepackage{subfiles} 

\begin{document}
\title{Topological phases in two-dimensional Rydberg lattices}
\author{Mathias B. M. Svendsen}
\author{Beatriz Olmos}
\affiliation{Institut f\"ur Theoretische Physik, Universit\"at Tübingen, Auf der Morgenstelle 14, 72076 T\"ubingen, Germany}

\begin{abstract}
    Rydberg lattice gases are at the forefront of quantum simulation platforms due to their inherent strong dipole-dipole interactions, long life-times and high degree of control currently achievable in experiments. We propose a simple and experimentally realizable two-dimensional lattice model consisting of two offset square sublattices of Rydberg atoms for the simulation of a so-called two-dimensional SSH model. This model reveals a plethora of topological phases, all connected through the variation of the lattice geometry. We predict the appearance of edge and corner states characterizing topological insulators, as well as pairs of topologically charged Dirac points associated with tilted and anisotropic cones in gapless semimetallic phases. The position, tilt and degree of anisotropy of the cones, which may determine the transport porperties of the system, are tunable via the sublattice offset parameters.
\end{abstract}

\maketitle

\textit{Introduction.---}Topological insulators and superconductors are systems characterized by a gapped bulk energy spectrum and by the presence of gapless edge or surface states \cite{Hasan2010,Qi2011}. Such systems may display, moreover, interesting features such as flat energy bands and a variety of quantum Hall effects \cite{Wen1995,Tang2011,Chang2023,He2024}. On the other hand, gapless topological systems, such as Weyl, Dirac and nodal line semimetals, are characterized by a gapless spectrum and the presence of topologically charged points, lines or surfaces, which may be exploited for their potential for controllable topological charge transport \cite{Burkov2016,Armitage2018,Rui2018,Hu2019}. The properties of a system in any of these topological phases are determined by the system's symmetries, making them robust against local perturbations such as disorder. This robustness is one of the main driving forces for the huge interest on the realization of such phases in a variety of experimental platforms, ranging from two- and three-dimensional materials to ultracold gases and photonic systems, among others \cite{Cooper2019,Wintersperger2020,Liang2022,Lu2014,Ozawa2019,deGroot2022}.





Among the many models that possess topological properties, the two-dimensional (2D) extension of the well known one-dimensional Su–Schrieffer–Heeger (SSH) model \cite{Su1979} has recently attracted much attention \cite{vanNiekerk2024,Luo2023,Gone2024,Agrawal2023,Min2024}. The reason may be found in its versatility: the 2D SSH model allows for the exploration of a broad range of topological insulators, including higher order topological insulators (HOTI), which exhibit no gapless edge states but rather topologically protected hinge or corner states \cite{Wu2020,Lui2019,Yang2024}. Moreover, some versions of this model lead to rich gapless topological physics such as the existence of Dirac cones, nodal lines and semimetallic phases \cite{Li2022,Li2022b,Jeon2022,Duan2024}. Driven by these interesting features, several platforms have been proposed for the realization of variants of the 2D SSH model, including photonic crystals \cite{Liu2018,Xie2019,Kim2020,Yan2023}, acoustic networks \cite{Huo2021,Zheng2019} and sphere-rod structures \cite{Liu2024}. Rydberg atoms have already proven to be a promising platform for the study topological features due to their high controllability, long lifetimes and strong dipolar interactions \cite{Browaeys2020}. Symmetry-protected topological phases and topological edge states have been theoretically proposed and experimentally observed in Rydberg lattices \cite{Weber2018,Deleseleuc2019,Li2021} and in Rydberg systems with synthetic dimensions \cite{Kiffner2017,Kanungo2022}. Moreover, topological pumping in Rydberg chains have been recently proposed \cite{Svendsen2024,Trautmann2024}. 

In this paper, we demonstrate the presence of a broad range of topological phases in a simple and experimentally realizable 2D SSH model consisting of Rydberg atoms. We find a semimetallic phase characterized by a pair of oppositely topologically charged and tuneable Dirac points and associated tilted anisotropic cones, which, for specific parameters, collapse into nodal lines. Furthermore, we identify and characterize three distinct gapped topological phases. Due to the coexistence of time-reversal and inversion symmetries, the Berry curvature vanishes throughout the first Brillouin zone, resulting in a zero Chern number. Interestingly, non-trivial topological phases can still be found in the system, which can be attributed to another topological invariant, the so-called 2D Zak phase \cite{Liu2018,Liu2017,Pratama2024}. All of these phases can be accessed by tuning the relative offset between two square Rydberg lattices, making this system the ideal platform for the experimental study of a broad spectrum of topological phases.

\begin{figure}[ht!]
    \centering
    \includegraphics[width =\columnwidth]{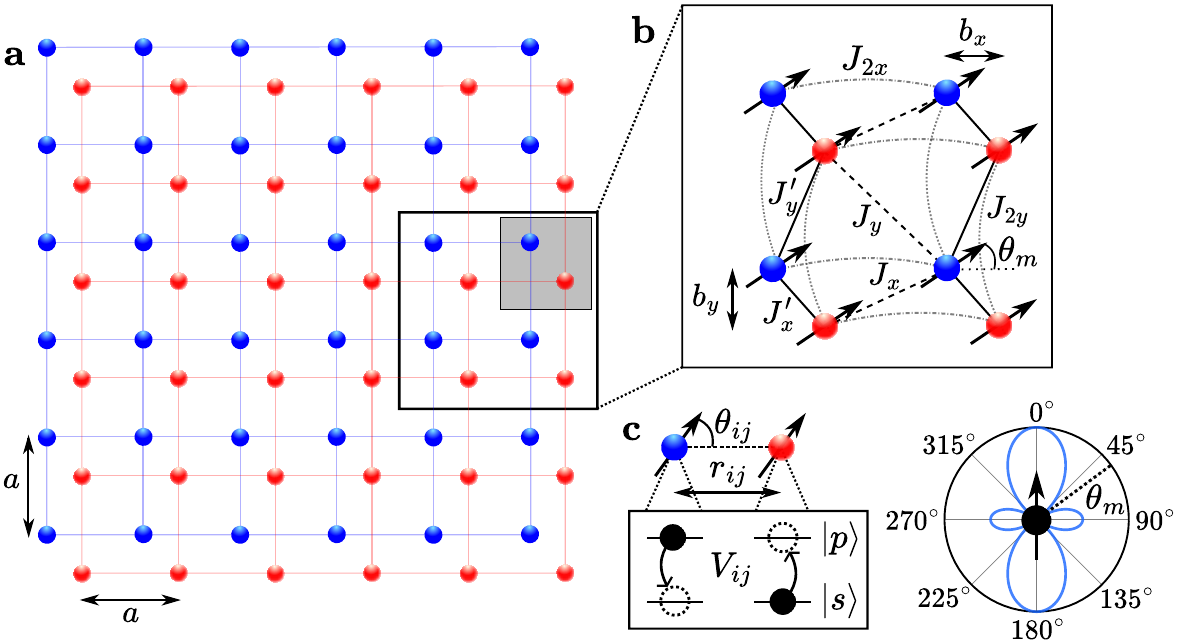}
    \caption{\textit{2D SSH model with Rydberg atoms.} \textbf{a}: We consider two interspersed square sublattices $A$ and $B$ represented by blue and red dots, with lattice constant $a$. \textbf{b:} Sublattice $B$ is offset with respect to sublattice $A$ by $0\leq b_x\leq a$ and $0\leq b_y\leq a$ in the $x$ and $y$-direction, respectively. Excitation exchange can occur between each site on one sublattice to the four nearest neighbors of the other sublattice with hopping parameters $J_x'$, $J_x$, $J_y'$ and $J_y$, and within the same sublattice with hopping parameters $J_{2x}$ and $J_{2y}$. \textbf{c:} Each lattice site contains an atom modeled as a two-level system where both levels are Rydberg levels denoted as $\ket{s}$ and $\ket{p}$. The direction of the transition dipole moment for each atom is indicated by the black arrows in \textbf{b}. The dipoles are tilted with an angle $\theta_m$ with respect to the $x$-axis. If $\theta_m=\arccos{1/\sqrt{3}}$, the couplings between the same sublattice in the $x$-direction become zero, i.e., $J_{2x}=0$.}
    \label{fig:fig1}
\end{figure}

\textit{Realization of the 2D SSH model with Rydberg atoms.---}We consider a two-dimensional system formed by two square lattices with $M\times M$ sites and lattice constant $a$ each. We denote the lattices as sublattice $A$ and $B$, depicted in Figure \ref{fig:fig1}a in blue and red, respectively. Sublattice $B$ is offset with respect to sublattice $A$ by a distance $b_x$ and $b_y$ in the $x$ and $y$-direction respectively, as shown in Fig. \ref{fig:fig1}b. Each site is populated by an atom, which we model as a two-level system. The two atomic electronic states are considered to be highly excited (Rydberg) levels. The resonant dipolar interactions between the two levels, namely $\ket{s}$ and $\ket{p}$ (see Fig. \ref{fig:fig1}c), is described by the Hamiltonian
\begin{equation}
    \hat{H}=\hbar \sum_{i\neq j}V_{ij}\hat{\sigma}_i^\dag\hat{\sigma}_j,
\end{equation}
where we have introduced the spin ladder operators ${\hat{\sigma}_j=\ket{s_j}\!\bra{p_j}}$ and $\hat{\sigma}_j^\dag=\ket{p_j}\!\bra{s_j}$. The dipole-dipole coupling, or excitation exchange rate between two atoms is given by
\begin{equation}\label{eq:V}
        V_{ij} = \frac{|\mathbf{d}|^2}{4\pi\epsilon_0\hbar} \frac{3\cos^2{\theta_{ij}}-1}{r_{ij}^3},
\end{equation}
where $r_{ij}$ is the distance between between the atoms and $\mathbf{d}$ is the transition dipole moment between the Rydberg states. Importantly, the coupling depends on the angle $\theta_{ij}$ between the dipoles and the axis connecting the atoms. We assume this dipole moment to be aligned along the angle $\theta_m$ with respect to the $x$-axis for all atoms, as illustrated in Figure \ref{fig:fig1}c. Due to the rapid decay of the coupling with the distance $r_{ij}$, it is a good approximation to assume that from any given site from one sublattice an excitation can hop only to any of the neighboring sites on the other sublattice, with rates denoted as $J_x$, $J_x'$, $J_y$, $J_y'$ obtained by evaluating Eq. \eqref{eq:V}. We will also consider a weak intrasublattice hopping in the $x$ and $y$-direction given by the hopping parameters $J_{2x}$ and $J_{2y}$ respectively (see Figure \ref{fig:fig1}b and the Supplemental Material \cite{supmat} for details). Note, however, that all the results here remain qualitatively unchanged when considering long-ranged coupling between distant atoms. 


It is easy to map this system into a two-dimensional SSH model by considering it as a square lattice formed by $M\times M$ two-site unit cells, each hosting one site of sublattice $A$ and one site of sublattice $B$. The system can then be seen as a stacking of 1D SSH chains either in the $x$- or in the $y$-direction. Introducing the annihilation and creation operators of an excitation (i.e. a hard-core boson) in sublattice $A$ and $B$ in cell $(n,m)$, with $n$ and $m$ running from 1 to $M$, such that $\hat{\sigma}_{M(n-1)+m}=\hat{a}_{nm}$ and $\hat{\sigma}_{Mn+m}=\hat{b}_{nm}$, the Hamiltonian can be rewritten as
\begin{equation}
\begin{split}
    \hat{H} &= \hbar J_x'\!\!\sum_{n,m=1}^{M}\hat{a}_{nm}^\dagger \hat{b}_{nm}+\hbar J_x\!\!\sum_{n=1}^{M-1}\!\sum_{m=1}^{M}\hat{b}_{nm}^\dagger \hat{a}_{n+1 m} \\
    &+\hbar J_y'\!\!\sum_{n=1}^{M}\!\sum_{m=1}^{M-1}\hat{a}_{nm}^\dagger \hat{b}_{nm+1}+\hbar J_y\!\!\!\sum_{n,m=1}^{M-1}\hat{b}_{nm+1}^\dagger \hat{a}_{n+1 m}\\ &+\hbar J_{2x}\!\!\sum_{n=1}^{M-1}\!\sum_{m=1}^{M}(\hat{a}_{nm}^\dagger \hat{a}_{n+1m}+\hat{b}_{nm}^\dagger \hat{b}_{n+1m})\\
    &+\hbar J_{2y}\!\!\sum_{n=1}^{M}\!\sum_{m=1}^{M-1}(\hat{a}_{nm}^\dagger \hat{a}_{nm+1}+\hat{b}_{nm}^\dagger \hat{b}_{nm+1}) + \text{h.c.}.
\end{split}
    \label{eq:HamiltonianRealSpace}
\end{equation}
In this work, we are interested in the properties of the system in the regime when at most one Rydberg excitation $\ket{p}$ is present in the whole system. In this limit, and considering periodic boundary conditions, the Hamiltonian \eqref{eq:HamiltonianRealSpace} can be diagonalized via a Fourier transform, and it may be written in momentum space as 
\begin{equation}
    \hat{H} = \sum_{\substack{\mathbf{k}\in \text{2D FBZ}\\ \alpha,\beta\in A,B}} \hat{c}^\dag_{\alpha}(\mathbf{k}) h_{\alpha\beta}(\mathbf{k}) \hat{c}_{\beta}(\mathbf{k}), 
\end{equation}
where $\hat{c}_\alpha(\mathbf{k}) = \hat{a}_{\mathbf{k}}$ for $\alpha = A$ and $\hat{c}_\alpha(\mathbf{k}) = \hat{b}_{\mathbf{k}}$ for $\alpha = B$, where $\hat{a}_{\mathbf{k}}$ and $\hat{b}_{\mathbf{k}}$ are the Fourier-transformed annihilation operator on sublattice $A$  and $B$, respectively. This momentum space Hamiltonian is given by
\begin{equation}
    h(\mathbf{k}) =
    \begin{pmatrix}
        n_0(\mathbf{k}) && n(\mathbf{k}) \\ 
        n^*(\mathbf{k}) && n_0(\mathbf{k})
    \end{pmatrix},
    \label{eq:BulkHamiltonian}
\end{equation}
where
\begin{equation}
    n(\mathbf{k}) \!=\! \hbar\left[ J_x'\!+\!J_y'e^{ik_ya}\!+\!J_xe^{-ik_xa}\!+\!J_ye^{-i(k_x-k_y)a}\right],
\end{equation}
and 
\begin{equation}\label{eq:n0}
    n_0(\mathbf{k}) = 2\hbar\left[J_{2x}\cos{(k_xa)}+J_{2y}\cos{(k_ya)}\right].
\end{equation}
These two functions determine both the topological properties and the single-particle excitation spectrum of the 2D SSH Hamiltonian \eqref{eq:HamiltonianRealSpace}.

\begin{figure*}[t]
    \centering
    \includegraphics[width =\linewidth]{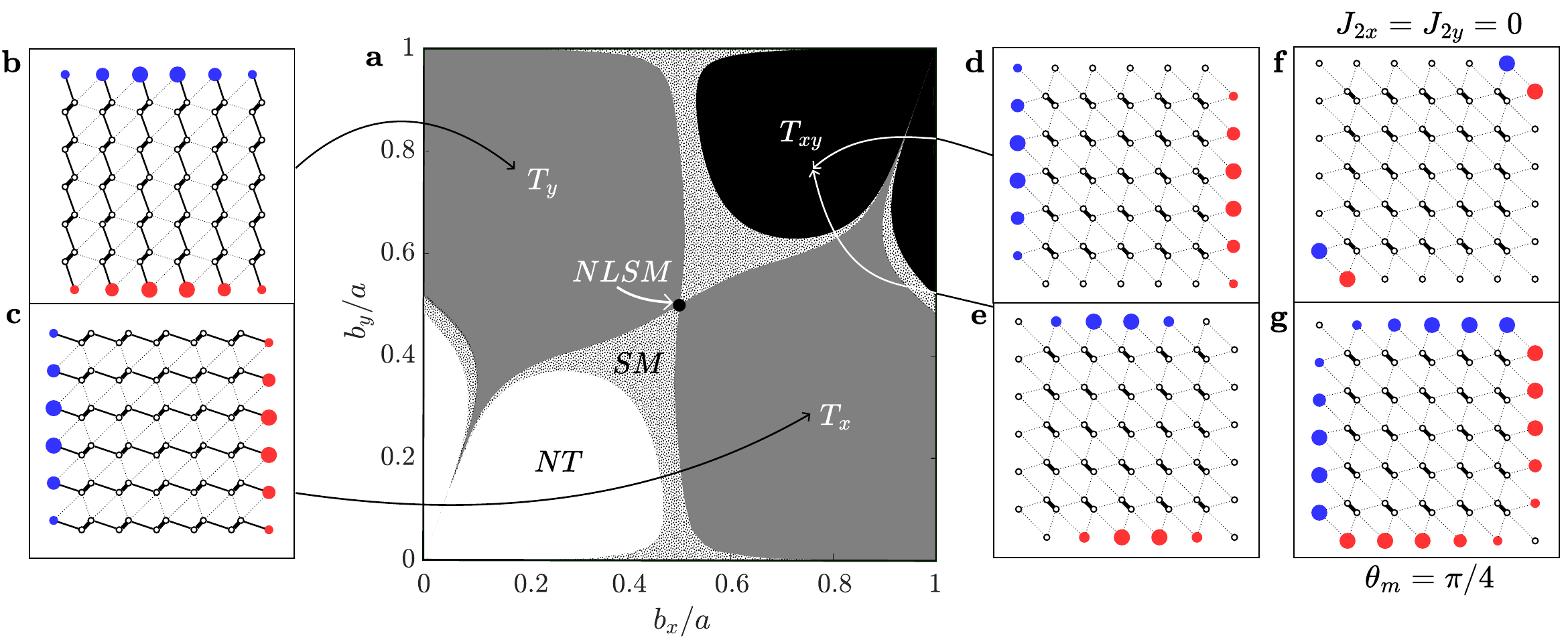}
    \caption{\textit{Phase diagram and edge states.} \textbf{a:} Phase diagram obtained fixing $\theta_m=\arccos{1/\sqrt{3}}$ and varying the sublattice offsets $b_x$ and $b_y$. Parameter spaces where Dirac semimetals and a nodal line semimetal appear are denoted by $SM$ and $NLSM$, respectively. In addition, four gapped phases can be found: A non-topological ($NT$) phase and three topological insulators, namely $T_x$ when $\mathbf{Z}=(\pi,0)$, $T_y$ when $\mathbf{Z}=(0,-\pi)$, and $T_{xy}$ when $\mathbf{Z}=(\pi,-\pi)$. \textbf{b-e:} Representative edge states for a finite system of $2\times M\times M=72$ atoms at the points \textbf{b:} $b_x=a/4$ and $b_y=3a/4$, \textbf{c:} $b_x=3a/4$ and $b_y=a/4$, \textbf{d} and \textbf{e:} $b_x=3a/4$ and $b_y=3a/4$. \textbf{f:} Corner states found at the same point, $b_x=3a/4$ and $b_y=3a/4$, but setting $J_{2x}=J_{2y}=0$. \textbf{g:} These corner states become polynomially localized at the same point in parameter space but with $\theta_m=\pi/4$, such that $J_{2x}=J_{2y}\neq0$.}
    \label{fig:PhaseDiagram}
\end{figure*}

Let us start by analyzing the symmetries of the system. The momentum space Hamiltonian \eqref{eq:BulkHamiltonian} exhibits both time-reversal symmetry, i.e. $\mathcal{T}h(\mathbf{k})\mathcal{T}^{-1} = h(-\mathbf{k})$ with $\mathcal{T}$ being the conjugation operation, and inversion symmetry, i.e. $\mathcal{I}h(\mathbf{k})\mathcal{I}^{-1} = h(-\mathbf{k})$ with $\mathcal{I}=\sigma_x$,
while chiral symmetry is in general broken due to the presence of the diagonal term $n_0(\mathbf{k})$, i.e. $\mathcal{S}h(\mathbf{k})\mathcal{S}^{-1} \neq -h(\mathbf{k})$ with $\mathcal{S}=\sigma_z$.
Independently of whether the chiral symmetry is broken or not, due to the coexistence of both time-reversal and inversion symmetry, the Berry curvature is zero throughout the full first Brillouin zone (FBZ), resulting in a zero Chern number. However, the topological gapped phases in this 2D SSH model can be characterized by another $\mathbb{Z}_2$ topological invariant, the so-called 2D Zak phase \cite{Resta1994,Liu2017}, defined as the vector
\begin{equation}
    \mathbf{Z}= (Z_x,Z_y) = \int\!\!\int_\mathrm{FBZ} dk_xdk_y \langle u(\mathbf{k})|i\partial_{\mathbf{k}}|u(\mathbf{k})\rangle,
\end{equation}
where $|u(\mathbf{k})\rangle$ is the eigenstate of $h(\mathbf{k})$ associated with one of the bands (here we use the lower band). For the 2D Zak phase to change its value, the parameters of the system need to be modified such that the gap between the energy bands closes in one of the high-symmetry points of the lattice. Hence, in order to find the boundaries between different topological phases, we calculate the single excitation spectrum of the system $E_\pm(\mathbf{k})=n_0(\mathbf{k})\pm|n(\mathbf{k})|$, and find the parameter values at which the gap between the two bands closes, i.e., $n(\mathbf{k})=0$ (see \cite{supmat} for details).

We fix now the value of the angle $\theta_m=\arccos{(1/\sqrt{3})}$ of the dipoles with respect to the $x$-axis, such that $J_{2x}=0$. By varying the values of the offset distances $b_x$ and $b_y$, we arrive at the phase diagram for the 2D SSH model with Rydberg atoms depicted in Figure \ref{fig:PhaseDiagram}a. Note that, varying the angle $\theta_m$ in general only modifies the exact position of the phase boundaries and hence the found phases remain robust against small variations of $\theta_m$. In the following, we will analyze in detail each one of the distinct gapped and gapless phases that arise in each parameter regime.

\begin{figure*}[t]
    \centering
    \includegraphics[width =\linewidth]{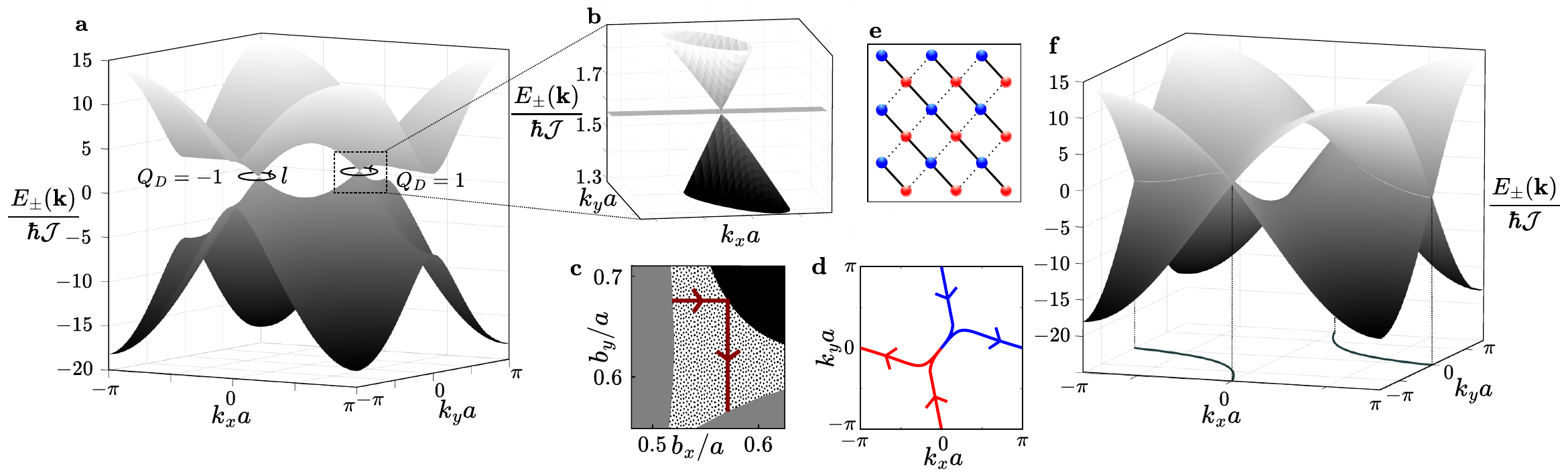}
    \caption{\textit{Tilted Dirac cones.} \textbf{a:} Energy bands $E_\pm(\mathbf{k})$ [normalized by $\hbar{\cal J}= |\mathbf{d}|^2/(4\pi \epsilon_0a^3)$] for $b_x=b_y=0.6a$. The semimetal is characterized by a pair of Dirac points with opposite topological charge $Q_D = \pm 1$. $l$ indicates the integration loop (see main text). \textbf{b:} The Dirac cones are both anisotropic and tilted. \textbf{c:} Trajectory in $b_x$-$b_y$ parameter space within the SM zone and \textbf{d:} corresponding trajectories of the pair pf Dirac cones withing the FBZ (blue and red represent $Q_D=+1$ and $-1$, respectively). \textbf{e:} At $b_x=b_y=a/2$ the system becomes mirror symmetric. \textbf{f:} At this point, the two Dirac cones become nodal lines, whose positions within the FBZ are projected on the $E(\mathbf{k})=-25\hbar{\cal J}$ plane.}
    \label{fig:TiltedDiracCones}
\end{figure*}

\textit{Topological gapped phases.---}Let us start by analyzing the three distinct gapped topologically non-trivial phases that one can explore in this system. Each of them has associated a non-zero 2D Zak phase, namely phase $T_x$ when $\mathbf{Z}=(\pi,0)$, $T_y$ when $\mathbf{Z}=(0,-\pi)$, and $T_{xy}$ when $\mathbf{Z}=(\pi,-\pi)$ (see Fig. \ref{fig:PhaseDiagram}a). The $T_x$ and $T_y$ phases are instances of weak topological insulators (WTIs) \cite{Ringel2012,Duan2024}, and are the result of stacking lower-dimensional topological insulators (here one-dimensional SSH chains). They are characterized by featuring edge states which only occupy certain edges of the insulator. Contrary to what the name suggests, WTIs actually display a remarkable topological protection against disorder \cite{Ringel2012}. In Figures \ref{fig:PhaseDiagram}b and c we show examples of an edge state for the $T_y$ and $T_x$ phases, respectively. These are obtained from the exact diagonalization of the Hamiltonian \eqref{eq:HamiltonianRealSpace} for a small system with $M=6$ unit cells in each direction. In the $T_y$ phase, which we can regard as a horizontal stacking of vertical SSH chains which interact weakly with one another, edge states are exponentially localized only along the boundaries of the lattice in the $y$-direction. Conversely, the $T_x$ phase can be seen as a vertical stacking of horizontal SSH chains. Correspondingly, here the edge states occupy the boundaries in the $x$-direction. Note, that strictly speaking the edge states in $T_y$ (Figure \ref{fig:PhaseDiagram}b) have a non-zero energy due to the non-zero value of the intrasublattice coupling $J_{2y}\neq 0$. However, as it can be found in the Supplemental material \cite{supmat}, these edge states are still extremely well separated from the bulk ones.

In the limit $J_{2y}=J_{2x} = 0$ (Fig. \ref{fig:PhaseDiagram}f), the last topological gapped phase, $T_{xy}$, becomes a higher order topological insulator (HOTI), characterized by edge states exponentially localized on the boundaries of the boundaries, which in two dimensions are the corners of the lattice. These corner states are topologically protected by the coexistence of time-reversal, particle-hole and chiral symmetries \cite{Schindler2018,Ghosh2024}. In the physical platform we consider here, though, this limit cannot be achieved by varying the parameters of the system. As indicated above, for $\theta_m = \arccos{(1/\sqrt{3})}$, $J_{2x}=0$, but $J_{2y}\neq0$. This in turn gives rise to two subsets of edge states in the $T_{xy}$ phase (see Figs. \ref{fig:PhaseDiagram}d and e): One subset has zero energy and the population is localized in the $y$-boundaries, while in the second one is localized in the $x$-boundaries but has a non-zero energy. On the other hand, by fixing the value of the angle $\theta_m=\pi/4$ such that $J_{2y} = J_{2x}$, the two sets of edge states acquire the same non-zero energy, which in turn gives rise to corner states, which are polynomially localized in the corners of the finite system (see Fig. \ref{fig:PhaseDiagram}g).

\textit{Tunable Dirac points and nodal lines---}In the phase diagram in Fig. \ref{fig:PhaseDiagram}a, we have also identified parameter regimes where the bandgap remains closed. Here, the system is in a semimetallic phase characterized by the emergence of a pair of Dirac points, which are singular touching points in the FBZ (see Fig. \ref{fig:TiltedDiracCones}a). Each Dirac point carries a topological charge given by
\begin{equation}
    Q_D = \frac{1}{2\pi i}\oint_l d\mathbf{k}\cdot n(\mathbf{k})^{-1}\partial_{\mathbf{k}} n(\mathbf{k}), 
\end{equation}
where $l$ is a path encircling the Dirac point \cite{Li2022b,Schnyder2008,Schnyder2011,Heikkila2011}. The two Dirac points have quantized opposite charges (or chirality) of $\pm 1$. These pairs of Dirac cones show a remarkable robustness, since they can only be annihilated by merging in one of the high-symmetry points of the FBZ. 

As one can observe in Figure \ref{fig:TiltedDiracCones}b, the dispersion relation around each Dirac point is linear, but in general anisotropic. Moreover, here we observe that the Dirac cones are not only anisotropic, but also tilted \cite{Amo2019,Kawarabayashi2011,Cheng2017,Lang2023}. This tilt comes as a result of the breaking of chiral symmetry stemming from the $J_{2y}\neq 0$ coupling. Anisotropic and tilted Dirac cones have attracted attention recently due to their impact on a system's quantum (direction-dependent) transport properties \cite{Trescher2015}. They have been realized, for example, in graphene perturbed by external modulations or strain, and intrinsically in hydrogenated graphene \cite{Lu2016,Choi2010,Park2008}. In our system, the degree of anisotropy, tilt and exact position of the cones in the FBZ can all be tuned by changing the sublattice offsets $b_x$ and $b_y$ and angle $\theta_m$ (see Figs. \ref{fig:TiltedDiracCones}c and d for an illustrative trajectory of the two Dirac cones). 
The merging of the pair of Dirac points (boundaries of the semimetallic phase) mark a transition into a topological non-trivial phase or a trivial phase depending on the exact values of the sublattice offsets. At the exact merging point, the Dirac cones display a linear dispersion relation in one direction and a parabolic dispersion relation in the orthogonal direction, similar to what has been found, e.g., in honeycomb lattices \cite{Tarruell2012,Montambaux2009} and in a variant of the 2D SSH model \cite{Li2022b}.

Finally, a unique instance of the semimetallic gapless phase occurs when $b_x = b_y = a/2$. Here, the system becomes mirror symmetric with respect to the $y=\pm x$ lines (see Fig. \ref{fig:TiltedDiracCones}e), i.e. $J_x=J_y'$ and $J_x'=J_y$. As a consequence, the Dirac cones collapse into nodal lines, as can be observed in Figure \ref{fig:TiltedDiracCones}f. Note, however, that this nodal-line semimetallic (NLSM) phase is not robust since any small changes in the parameters of the system lead to the breaking of the mirror symmetry and hence either the nodal lines become Dirac cones or the system becomes gapped and enters the WTI phases $T_x$ or $T_y$ (see Figure \ref{fig:PhaseDiagram}a).

\textit{Conclusion and outlook.---}We have established theoretically the suitability of a Rydberg lattice gas as an ideal platform for the exploration of topological phenomena. The uniqueness of this model lies in its tunability: simply changing the offset between the lattices and the dipole orientation allows to explore a broad range of distinct topological phases. However, in this work we have only started to tap into the full potential of the system by studying its single-excitation properties. The exploration of quantum many-body phenomena beyond the linear optics regime will be our next research direction. Particularly interesting will be to witness an experimental implementation of the proposed system, which is at reach with currently available arrays of optical tweezers setups \cite{Deleseleuc2019,Hollerith2022,Steinert2023}. For illustration purposes, one can consider ${}^{87}\text{Rb}$ atoms individually trapped in two arrays of optical tweezers with lattice parameter $a=12\,\mu$m. When the two Rydberg levels are chosen to be the $\ket{60S_{1/2}}$ and the $\ket{60P_{1/2}}$ state, all couplings described in this paper are on the order of MHz, with bandgaps that vary between 10 and 100 MHz.

\acknowledgments
\textit{Acknowledgments.--}The authors thank Prof. Christian Groß and Dr. Björn Sbierski for fruitful discussions. We acknowledge funding from the Deutsche Forschungsgemeinschaft within the Grant No. 452935230 and the research unit FOR5413 (Grant No. 465199066).

\onecolumngrid
\clearpage
\subfile{Supplementary.tex}


\end{document}

%% file: Supplementary.tex
\onecolumngrid
\makeatletter
\renewcommand{\theequation}{S\arabic{equation}}
\renewcommand{\thefigure}{S\arabic{figure}}
\renewcommand{\thetable}{S\arabic{table}}
\setcounter{secnumdepth}{1}

\begin{center}
{\Large SUPPLEMENTAL MATERIAL}
\end{center}
\begin{center}
\vspace{0.8cm}
{\Large Topological phases in two-dimensional Rydberg lattices}
\end{center}
\begin{center}
Mathias B. M. Svendsen$^1$ and Beatriz Olmos$^1$
\end{center}
\begin{center}
$^1${\em Institut f\"ur Theoretische Physik and Center for Integrated Quantum Science and Technology, Universit\"at T\"ubingen, Auf der Morgenstelle 14, 72076 T\"ubingen, Germany}\\
\end{center}

In this supplemental material we detail the geometry of the 2D Rydberg lattice and how the parameters of the Hamiltonian relate to the geometry of the lattice. We briefly discuss the bulk properties of the system and show that the bulk energy gap closes for four different combinations of the hopping rates. We discuss edge states in a semi-infinite system by making a partial Fourier transform of the Hamiltonian. We show that the edge states are not at zero energy, due to the broken chiral symmetry, but at some finite energy still well separated from the bulk energies. 
\section{Geometry of the Rydberg lattice and bulk properties}

\begin{figure}[h]
    \centering
    \includegraphics[width =0.3\columnwidth]{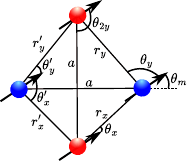}
    \caption{\textit{Geometry of the Rydberg lattice.} The distances and angles between the Rydberg atoms in the lattice.}
    \label{RydbergAngles}
\end{figure}
\begin{figure}[t]
    \centering
    \includegraphics[width =0.75\columnwidth]{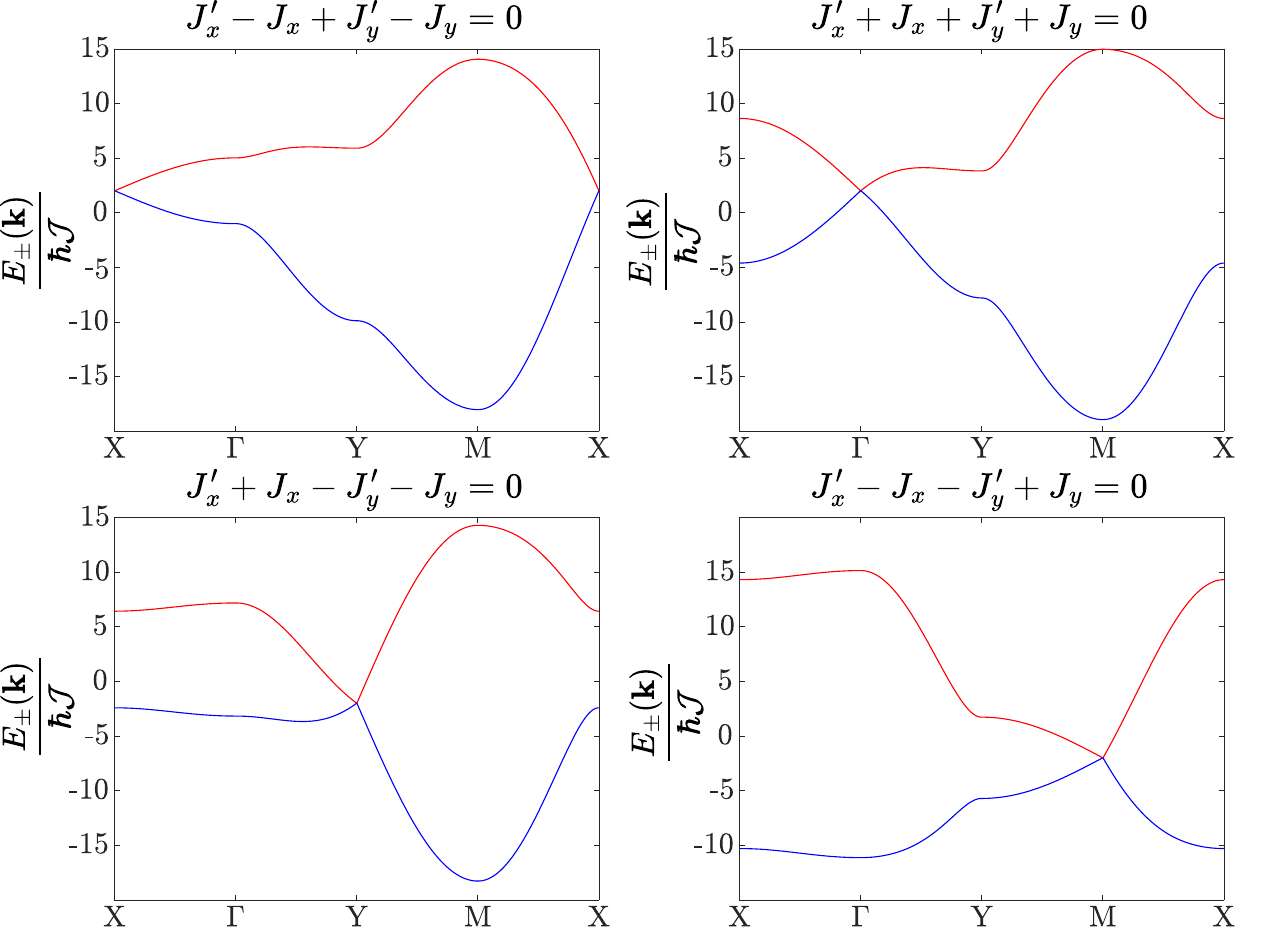}
    \caption{\textit{Bulk spectra.} The bulk energy gap closes at the high symmetry points (X, $\Gamma$, Y and M) in the FBZ for different values of the hopping parameters given above and for $\theta_m = \arccos{(1/\sqrt{3})}$. The bulk energy is normalized by the factor $\hbar \mathcal{J} = |\mathbf{d}|^2/(4\pi \epsilon_0a^3)$.}
    \label{BandsSymmetryPoints}
\end{figure}

The Hamiltonian of the 2D Rydberg lattice consisting of two $N\times M$ sublattices is given by 
\begin{equation}
\begin{split}
    \hat{H} &= \hbar J_x'\sum_{n=1}^{N}\sum_{m=1}^{M}\hat{a}_{nm}^\dagger \hat{b}_{nm}+\hbar J_x\sum_{n=1}^{N-1}\sum_{m=1}^{M}\hat{b}_{nm}^\dagger \hat{a}_{n+1 m}+\hbar J_y'\sum_{n=1}^{N}\sum_{m=1}^{M-1}\hat{a}_{nm}^\dagger \hat{b}_{nm+1}+\hbar J_y\sum_{n=1}^{N-1}\sum_{m=1}^{M-1}\hat{b}_{nm+1}^\dagger \hat{a}_{n+1 m}\\ &+\hbar J_{2x}\sum_{n=1}^{N-1}\sum_{m=1}^{M}(\hat{a}_{nm}^\dagger \hat{a}_{n+1m}+\hat{b}_{nm}^\dagger \hat{b}_{n+1m})+\hbar J_{2y}\sum_{n=1}^{N}\sum_{m=1}^{M-1}(\hat{a}_{nm}^\dagger \hat{a}_{nm+1}+\hat{b}_{nm}^\dagger \hat{b}_{nm+1}) + \text{h.c.},
\end{split}
    \label{sup:HamiltonianRealSpace}
\end{equation}
where the hopping parameters are given by the Rydberg dipole-dipole interactions  
\begin{equation*}
    \begin{split}
        J_x' &= \frac{|\mathbf{d}|^2}{4\pi\epsilon_0\hbar} \frac{3\cos^2{\theta_{x}'}-1}{r_{x}'^3}, \qquad
        J_x = \frac{|\mathbf{d}|^2}{4\pi\epsilon_0\hbar} \frac{3\cos^2{\theta_{x}}-1}{r_{x}^3},\\
        J_y' &= \frac{|\mathbf{d}|^2}{4\pi\epsilon_0\hbar} \frac{3\cos^2{\theta_{y}'}-1}{r_{y}'^3}, \qquad
        J_y = \frac{|\mathbf{d}|^2}{4\pi\epsilon_0\hbar} \frac{3\cos^2{\theta_{y}}-1}{r_{y}^3}, 
        \\ 
        J_{2x} &= \frac{|\mathbf{d}|^2}{4\pi\epsilon_0\hbar} \frac{3\cos^2{\theta_m}-1}{a^3}, \qquad
        J_{2y} = \frac{|\mathbf{d}|^2}{4\pi\epsilon_0\hbar} \frac{3\cos^2{\theta_{2y}'}-1}{a^3}.
    \end{split}
    \label{HoppingRydberg}
\end{equation*}
Here $r_x'= \sqrt{b_x^2+b_y^2}$, $r_x= \sqrt{(a-b_x)^2+b_y^2}$, $r_y'= \sqrt{b_x^2+(a-b_y)^2}$ and $r_y= \sqrt{(a-b_x)^2+(a-b_y)^2}$ (see Figure \ref{RydbergAngles}). The angles are given by 
\begin{equation*}
    \begin{split}
        \cos{\theta_x'} &= (b_x\cos{\theta_m}-b_y\sin{\theta_m})/r_x', \\  \cos{\theta_x} &= [(a-b_x)\cos{\theta_m}+b_y\sin{\theta_m}]/r_x, \\ \cos{\theta_y'} &= [b_x\cos{\theta_m}+(a-b_y)\sin{\theta_m}]/r_y', \\ \cos{\theta_y} &= [(a-b_x)\cos{\theta_m}-(a-b_y)\sin{\theta_m}]/r_y, \\
        \cos{\theta_{2y}} &= \cos{(\pi/2-\theta_m)}.
    \end{split}
\end{equation*}
Making a Fourier transform of the operators in the Hamiltonian \eqref{sup:HamiltonianRealSpace} as outlined in the main manuscript, the bulk properties of the system are found. The two energy bands of bulk system are given by $E_{\pm}(\mathbf{k}) = n_0(\mathbf{k}) \pm |n(\mathbf{k})|$. The energy gap, $E_+({\mathbf{k}})-E_-(\mathbf{k})$, closes for four different combinations of the hopping parameters in the four high symmetry points of the FBZ as shown in Figure \ref{BandsSymmetryPoints}.

\section{Semi-infinite system}

\begin{figure}[h]
    \centering
    \includegraphics[width =0.9\columnwidth]{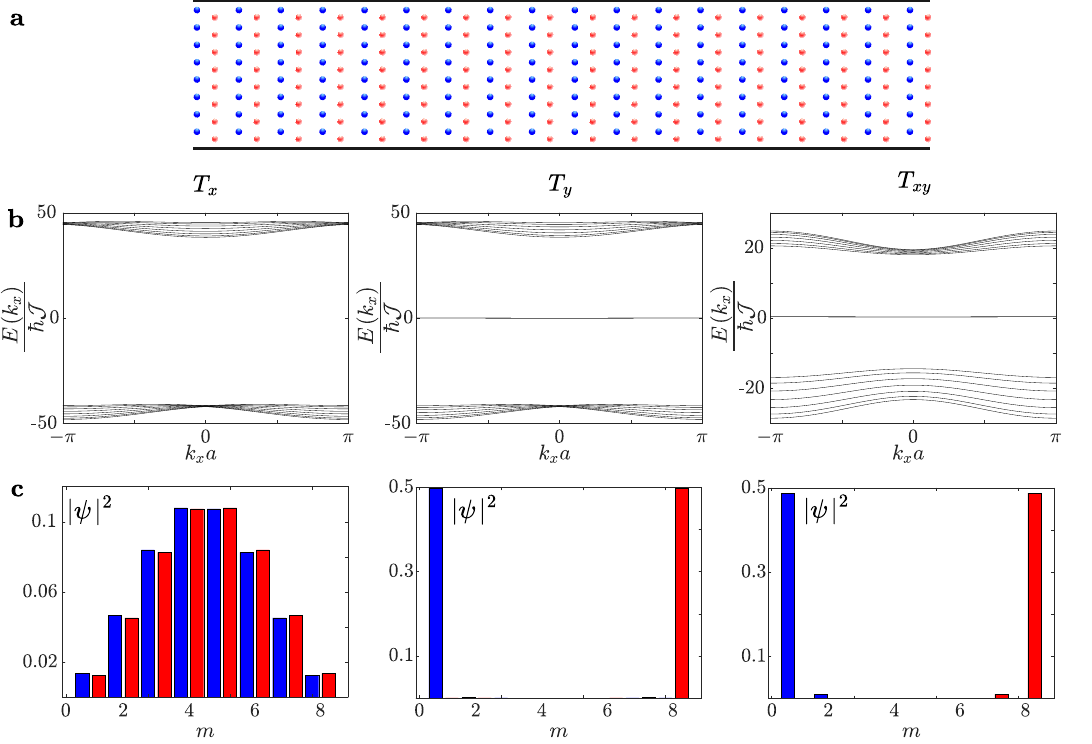}
    \caption{\textit{Semi-infinite system in the $x$-direction}. \textbf{a}: The sublattices are infinite in the $x$-direction and finite in the $y$-direction with $M=8$. The black lines indicates the boundary. \textbf{b}: The energies $E(k_x)$ of the partial Fourier transformed Hamiltonian $H(k_x)$ in the three topological gapped regions $T_x$, $T_y$ and $T_{xy}$. \textbf{c}: Magnitude of illustrative low-energy eigenstates, $|\psi|^2$ where blue/red bars correspond to the occupation of sites on sublattice $A/B$, respectively. The sublattice offsets are $b_x = 0.8a$ and $b_y=0.2a$ in the $T_x$ phase, $b_x = 0.2a$ and $b_y=0.8a$ in the $T_y$ phase and $b_x = 0.8a$ and $b_y=0.8a$ in the $T_{xy}$ phase.}
    \label{SemiInfX}
\end{figure}

To investigate the existence of edge states in a 2D system, it is useful to start by considering a semi-infinite system, i.e. keep one of the two dimensions small and the other infinite. For illustration purposes, we consider here two offset $N\times M$ sublattices, where $M$ is finite (here $M=8$) and $N$ is infinite. Doing so allows us to take a partial Fourier transform in the infinite $x$-direction of the operators in the  Hamiltonian given in equation \eqref{sup:HamiltonianRealSpace}, namely
\begin{equation*}
\begin{split}
    \hat{a}_{nm} &= \frac{1}{\sqrt{N}}\sum_{k_x} e^{ik_x n a}\hat{a}_{k_xm}, \\
    \hat{b}_{nm} &= \frac{1}{\sqrt{N}}\sum_{k_x} e^{ik_x n a}\hat{b}_{k_x m}.
\end{split}
\end{equation*}
The Hamiltonian may be written now in a simplified form $\hat{H} = \sum_{k_x} \hat{H}(k_x)$, where
\begin{equation*}
\begin{split}
        \hat{H}(k_x) &= \hbar\bigg\{\sum_{m=1}^{M} (J_x'+J_xe^{-ik_xa})\hat{a}_{k_x m}^\dagger \hat{b}_{k_x m} + (J_y'+J_ye^{ik_xa})\hat{b}_{k_x m+1}^\dagger \hat{a}_{k_x m} \\
        &+ J_{2y}(\hat{a}_{k_x m}^\dagger \hat{a}_{k_x m}+\hat{b}_{k_x m}^\dagger \hat{b}_{k_x m}) +\text{h.c.}\bigg\},
\end{split}
\end{equation*}
where $J_{2y}$ now acts as an on-site constant energy and $k_x$ acts as a parameter. Edge states can now be identified by diagonalizing the above Hamiltonian and inspecting the edge occupancy of the eigenstates. Figure \ref{SemiInfX} shows the eigenenergies $E$ as a function of $k_x$ in the three different gapped topological regions. The $T_y$ and $T_{xy}$ regions show an illustrative example of an edge state (which we call here $\psi$) at small but non-zero energy, well separated from the bulk energies, exponentially localized to both edges, while in the $T_{x}$ region only bulk states appear since the chains in the $x$-direction are infinite.

\begin{figure}[t]
    \centering
    \includegraphics[width =0.9\columnwidth]{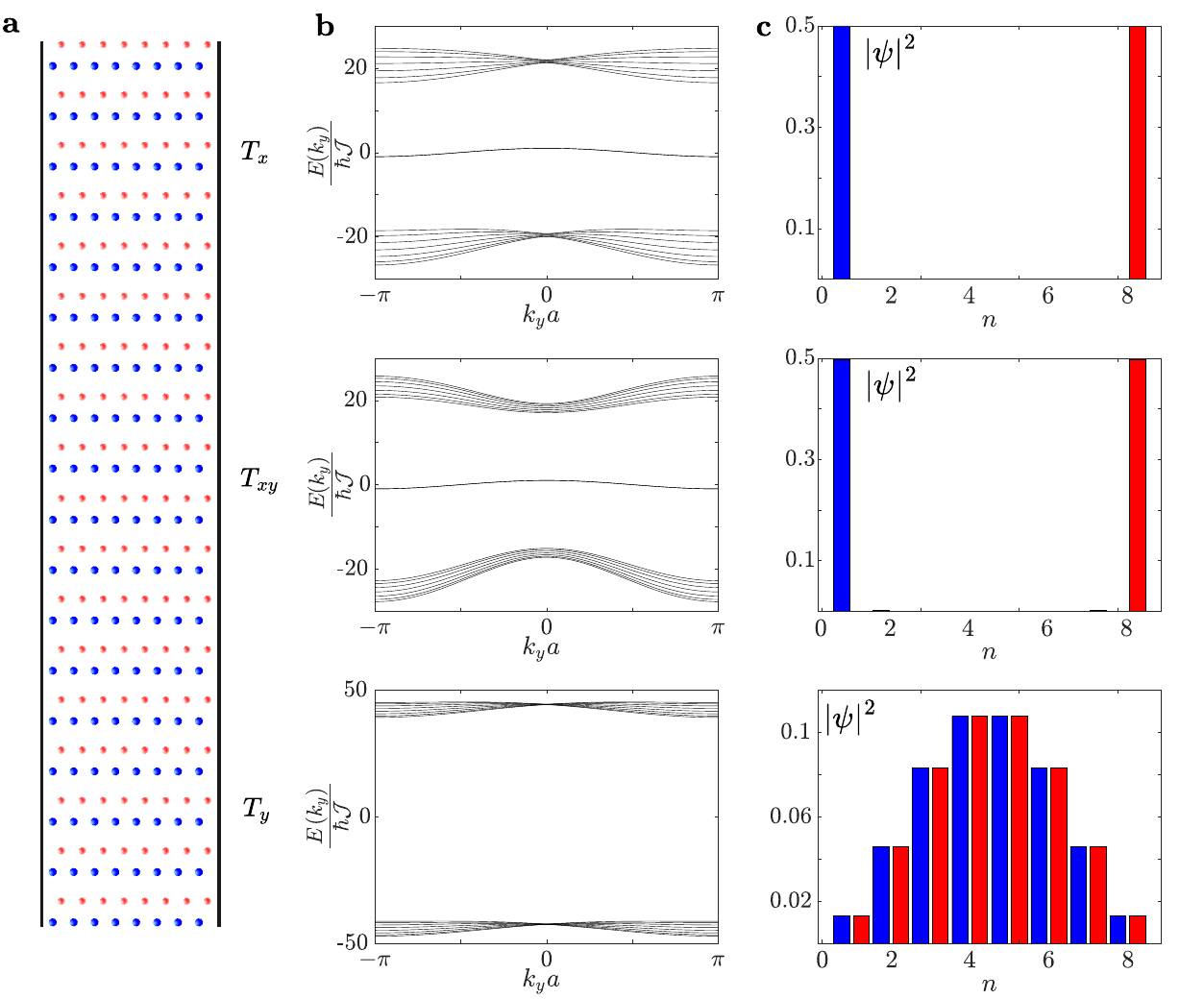}
    \caption{\textit{Semi-infinite system in the $y$-direction}. \textbf{a}: The sublattices are infinite in the $y$-direction and finite in the $x$-direction with $N=8$. The black lines indicates the boundary. \textbf{b}: The energies $E(k_y)$ of the partial Fourier transformed Hamiltonian $H(k_y)$ in the three topological gapped regions $T_x$, $T_y$ and $T_{xy}$. \textbf{c}: Magnitude of illustrative low-energy eigenstates, $|\psi|^2$ where blue/red bars correspond to the occupation of sites on sublattice $A/B$, respectively. The sublattice offsets are $b_x = 0.8a$ and $b_y=0.2a$ in the $T_x$ phase, $b_x = 0.2a$ and $b_y=0.8a$ in the $T_y$ phase and $b_x = 0.8a$ and $b_y=0.8a$ in the $T_{xy}$ phase.}
    \label{SemiInfY}
\end{figure}

The same partial Fourier transform can be performed when considering the opposite case, i.e. $M$ is infinite and $N=8$. Using a partial Fourier transform in the $y$-direction
\begin{equation*}
\begin{split}
       \hat{a}_{nm} &= \frac{1}{\sqrt{M}}\sum_{k_y} e^{ik_y m a}\hat{a}_{nk_y}, \\
    \hat{b}_{nm} &= \frac{1}{\sqrt{M}}\sum_{k_y} e^{ik_y m a}\hat{b}_{nk_y},
\end{split}
\end{equation*}
the Hamiltonian can be expressed now as $\hat{H} = \sum_{k_y}\hat{H}(k_y)$, with
\begin{equation*}
\begin{split}
    \hat{H}(k_y) &= \hbar\bigg\{\sum_{n=1}^{N} (J_x'+J_y'e^{ik_ya})\hat{a}_{nk_y}^\dagger \hat{b}_{nk_y} + (J_x+J_ye^{-ik_ya})\hat{b}_{nk_y}^\dagger \hat{a}_{n+1 k_y}  \\
    &+ J_{2y}e^{ik_ya}(\hat{a}_{n k_y}^\dagger \hat{a}_{n k_y}+\hat{b}_{n k_y}^\dagger \hat{b}_{n k_y})+ \text{h.c.}\bigg\}.
\end{split}
\end{equation*}
Equivalently to the previous case we can calculate the eigenstates of this Hamiltonian in the three topological gapped regions of the phase diagram (see Figure \ref{SemiInfY}). As expected, edge states appear in the $T_x$ and $T_{xy}$ regions while the eigenstates in the $T_y$ region are bulk states. Since the edge states occupy the boundaries in the $x$-direction, the intrasublattice interactions along this edge, $J_{2y}\neq0$, induce a non-zero dispersion relation for the edge state energies as seen in Figure \ref{SemiInfY}, i.e. the edge state energy $E(k_y)$ is not flat as a function of the momentum $k_y$.